\documentclass[lettersize,journal]{IEEEtran}
\usepackage{amsmath,amsfonts}
\usepackage{algorithmic}
\usepackage{array}
\usepackage[caption=false,font=normalsize,labelfont=sf,textfont=sf]{subfig}
\usepackage{textcomp}
\usepackage{stfloats}
\usepackage{url}
\usepackage{verbatim}
\usepackage{graphicx}
\usepackage{balance}

\usepackage{cite}
\usepackage{amsmath,amssymb,amsfonts}
\usepackage{algorithmic}
\usepackage{graphicx}
\usepackage{textcomp}
\usepackage{url}

\usepackage{bbm}
\usepackage{mathrsfs}
\usepackage{verbatim}
\usepackage{textcomp}
\usepackage{amsmath}
\usepackage{array}
\usepackage{booktabs}
\usepackage{longtable}
\usepackage{multirow}
\usepackage{lettrine}
\usepackage{nomencl}
\usepackage[mathscr]{euscript}
\usepackage{bm}
\usepackage{graphicx}  
\usepackage{float}  
\usepackage{makecell} 

\def\BibTeX{{\rm B\kern-.05em{\sc i\kern-.025em b}\kern-.08em
		T\kern-.1667em\lower.7ex\hbox{E}\kern-.125emX}}
\begin{document}
	 \title{Transferable Deep Learning Power System Short-Term Voltage Stability Assessment with Physics-Informed Topological Feature Engineering}
	
	
	\author{Zijian~Feng, Xin~Chen*, Zijian~Lv, Peiyuan~Sun, Kai~Wu,
	\thanks{This work was supported in part by the National Natural Science Foundation of China (Grant No.21773182 (B030103)) and the HPC Platform, Xi'an Jiaotong University.}
	\thanks{Zijian~Feng (e-mail: KenFeng@stu.xjtu.edu.cn),  Xin Chen (Corresponding author, e-mail: xin.chen.nj@xjtu.edu.cn), and Zijian~Lv (e-mail: jtlzj271828@stu.xjtu.edu.cn) are with the School of Electrical Engineering and the Center of Nanomaterials for Renewable Energy, State Key Laboratory of Electrical Insulation and Power Equipment, School of Electrical Engineering, Xi'an Jiaotong University, Xi'an, Shaanxi, China.}
	\thanks{Peiyuan~Sun (e-mail:spy2018@stu.xjtu.edu.cn) is with the School of Electrical Engineering, Xi'an Jiaotong University, Xi'an, Shaanxi, China.}
	\thanks{Kai~Wu (e-mail: wukai@mail.xjtu.edu.cn.) is with State Key Laboratory of Electrical Insulation and Power Equipment, School of Electrical Engineering, Xi'an Jiaotong University, Xi'an, Shaanxi, China.}
}
	
	\maketitle
	
	\begin{abstract}
Deep learning (DL) algorithms have been widely applied to short-term voltage stability (STVS) assessment in power systems. However, transferring the knowledge learned in one power grid to other power grids with topology changes is still a challenging task. This paper proposed a transferable DL-based model for STVS assessment by constructing the topology-aware voltage dynamic features from raw PMU data. Since the reactive power flow and grid topology are essential to voltage stability, the topology-aware and physics-informed voltage dynamic features are utilized to effectively represent the topological and temporal patterns from post-disturbance system dynamic trajectories. 
The proposed DL-based STVS assessment model is tested under random operating conditions on the New England 39-bus system. It has 99.99\% classification accuracy of the short-term voltage stability status using the topology-aware and physics-informed voltage dynamic features. In addition to high accuracy, the experiments show good adaptability to PMU errors. Moreover, The proposed STVS assessment method has outstanding performance on new grid topologies after fine-tuning. In particular, the highest accuracy reaches 99.68\% in evaluation, which demonstrates a good knowledge transfer ability of the proposed model for power grid topology change.

\end{abstract}


\section{Introduction}
\label{sec:introduction}

\IEEEPARstart{S}{hort-term} voltage stability (STVS) refers to the ability of bus voltages to recover rapidly to an acceptable level after a large disturbance\cite{BG1}. The leading cause of power system STVS is the fast-restoring dynamic load components, most typically induction motors (IMs), which would draw a high current in quick power demand recovery after the voltage drop\cite{BG3}. On the other hand, the low-voltage-ride-through capability is becoming more demanding with more inverter-interface renewable energy sources (RES) in modern power grids, leading to a more severe STVS problem. The blackout event in South Australia in 2016 reflected the vulnerability of STVS in power system and substantial economic losses resulting from short-term voltage instability\cite{BG4}. To avoid large-scale power outages and losses caused by short-term voltage instability, a real-time, accurate, and reliable STVS assessment model is necessary.

In the past, efforts have been devoted to accurate STVS assessment methods. However, the employment of various electronic components and the diversity and uncertainty of dynamic loads increase the complexity of the STVS problem. Due to the high computational burden, traditional methods, including time-domain simulation (TDS) and energy function methods, are not applicable for real-time stability assessment of the complex power system. 

With the development of advanced information and communication technologies, online monitoring system has been widely employed in power systems. The wide area measurement systems (WAMS) with phasor measurement units (PMUs) can capture the system-wide dynamic behavior of the power system and gain a large amount of high-resolution data \cite{BG5}. In recent years, numerous studies have been developed to realize reliable STVS assessment with data-driven methods owing to massive PMU data. A diversity of machine learning (ML) algorithms has been used in data-driven STVS assessment, including decision tree (DT), support vector machine (SVM), and extreme learning machine (ELM). DTs classify the system dynamic time series subsequences (shapelets) \cite{LR3, DT3} for STVS assessment, which is more interpretable for more straightforward rules. SVM model is utilized to extract the key features of anticipated faults of the power system and predict STVS status \cite{SVM1}. ELM and random vector functional link (RVFL) units are utilized to construct an ensemble-based randomized learning model for STVS assessment with snapshots as input features \cite{LR5,LR6}. Deep learning (DL) has recently attracted more attention since it can effectively extract features in data. Two DL algorithms, temporal convolutional neural network (TCN) and long short-term memory network (LSTM), are combined for real-time and accurate STVS assessment \cite{TCN}. Attention mechanisms are also employed to improve the feature extraction ability and capture the temporal dependence in time series by the BiGRU-attention-based model \cite{GAT}.

Meanwhile, with appropriate feature construction methods, the potential stability information in the time series can be fully represented in the form of feature matrices or heatmaps. These feature construction methods are an effective way to enhance the accuracy of dynamic stability assessment (DSA). Thus, various feature construction methods, wavelet transform (WT), symbolic Fourier approximation (SFA), and Gramian Angular Field (GAF) are used to covert the one-dimensional time series into two-dimensional feature images, and further cooperate with DL algorithms to realize real-time DSA \cite{LR13,LR14,GAF}.

However, the ignorance of topology information of the power grid is a common limitation in most of the existing data-driven STVS assessment methods. In a practical power grid, the potential topological correlation coupling among different buses would determine the trigger and spread of short-term voltage instability. Nevertheless, most of the studies still need to include these critical issues in constructing STVS assessment models. By incorporating geospatial information with shapelets, a spatial-temporal feature learning STVS model is established to enhance the effect of assessment \cite{LR15}. But it is not applicable to large-scale power grids due to its reliance on detailed geographic information of substations. Also, LSTM and Graph neural network (GNN) are employed to grasp spatial–temporal correlations behind voltage dynamics considering system structural and operational complexities \cite{LR21, GNN2}.

Moreover, another major limitation of the current data-driven online DSA arises from the requirement of applying the existing assessment models to a new power grid topology. Topological changes caused by significant disturbances, such as N-k (k $>1$) accidents, scheduled maintenance, etc., constitute the challenge for STVS assessment. Once there are any topology changes, the original data-driven model is not applicable. It is time-consuming to recollect the massive training dataset for rebuilding the model \cite{LR16}. Thus, inspired by transfer learning, an assessment model with good scalability that can perform in the different topologies of power systems is essential.

There are a few studies based on transfer learning in DSA. Transfer learning aims to propagate knowledge from the source domain to the target domain \cite{LR19}. Recently, some methods have been proposed to realize transferable transient stability assessment, but few focus on the STVS assessment. Orthogonal weight modification (OWN) and convolutional neural network (CNN) model are used to establish the model with continuous learning ability, which can update the model quickly and continuously \cite{LR17}. A deep belief network (DBN) is used to realize transfer learning between different operating conditions and topology \cite{LR18}. The dominant instability mode (DIM) identification models under base power grid topology are transferred to the N-1 topology scenarios \cite{LR22} through active transfer learning and unsupervised transfer learning. 

Aiming at the STVS assessment, this article proposes a DL-based power system STVS assessment model with topology-aware physics-informed feature engineering and knowledge transfer for power grid topology change. By extracting the physical features of the data, this paper aims to realize real-time STVS assessment at high accuracy. At the same time, this method extracts the power grid topological features related to STVS, which not only improves the model's performance but also helps to realize the transfer application among the power systems with different topologies. The main contributions of this article include the following:

\begin{enumerate}
\item The topology-aware and physics-informed voltage dynamic features are constructed from the raw PMU data, which are based on the grid topology and real-time reactive power demands on loads. The features contain both topological information and temporal information from voltage dynamics.
\item With the PMU dataset generated with the New England 39-bus system, the DL-based STVS assessment model shows highly accurate and robust performance with the topology-aware and physics-informed voltage dynamic features.
\item The topological-temporal information in voltage dynamic features enhances the knowledge transfer ability of the DL-based STVS assessment model for power grids with new topology. The proposed STVS assessment model demonstrates accurate STVS assessment with direct transfer learning and fine-tuning in the dataset of new power grids.

\end{enumerate}

\section{Topology-aware Dynamic Features}
Voltage stability is the ability that a power system maintains the voltages at all buses after being subjected to a disturbance. The topology of the power system is critical to voltage stability. The reactive power flow can determine the voltage adjustment. The reactive power $Q_{i}$ of node $i$,
\begin{eqnarray}
Q_{i}\triangleq-\sum_{j=1}^{n+m}V_{i}V_{j}B_{ij}
\end{eqnarray}
where $B_{ij}=b_{ij}\cos(\theta_{i}-\theta_{j})$, and $b_{ij}$ is the susceptance between nodes $i$ and $j$; $V_{i}$ and $\theta_{i}$ are the voltage magnitude and rotor angle of node $i$. 

The voltage stability in the power system is strongly associated with the reactive power flow since the driving force of instability is the power consumed by loads. The stress on loads by high reactive power consumption can cause voltage reduction. To produce a measure of stress on each load in a power system at steady state, a load matrix $\mathbf{L_{s}}$ \cite{FE1} can be defined as,
\begin{equation}
\mathbf{L_{s}}\triangleq \frac{1}{4} {\rm diag} (\mathbf{B_{LL}^{-1}}\mathbf{B_{LG}}\mathbf{V_{G}}) \mathbf{B_{LL}} {\rm diag} (\mathbf{B_{LL}^{-1}}\mathbf{B_{LG}}\mathbf{V_{G}})
\end{equation}
where $\rm diag(\cdot)$ is a matrix on the main diagonal. $\mathbf{B_{LL}}$ is the susceptance matrix among loads, $\mathbf{B_{LG}}$ is the susceptance matrix between loads and generators, $\mathbf{V_{G}}$ is the vector of generator voltages. Since steady-state voltage instability occurs over minutes and the generator swing dynamic occurs on the order of seconds, the generator voltages can be assumed as constant. $\mathbf{B_{LL}^{-1}}\mathbf{B_{LG}}\mathbf{V_{G}}$ is the open-circuit voltage on loads, when $\mathbf{V_{G}}$ stays unchanged. The load matrix $\mathbf{L_{s}}$ is based on the grid topology to indicate the sensitivity of the increasing stress on nodes. Moreover, to produce a node-by-node measure of grid stress and estimate the distance to voltage collapse, a voltage stability index \cite{FE1} $\Delta$ is defined as,
\begin{equation}
\Delta\triangleq \Vert \mathbf{L_{s}}^{-1}\mathbf{q_{L}}\Vert_{\infty},
\label{cond2}
\end{equation}
where $\mathbf{q_{L}}$ denotes reactive power demands of loads at steady state, which can be calculated as \cite{FE2},
\begin{equation}
\mathbf{q_{L}}=-[\mathbf{V_{L}}](\mathbf{B_{LL}}\mathbf{V_{L}}+\mathbf{B_{LG}}\mathbf{V_{G}}),
\end{equation}
where $[\mathbf{V_{L}}]$ is the diagonal matrix of load voltages.
With $\Delta$ as an indicator of steady-state voltage stability, there is a fixed bound that when $\Delta=1$, the network's guaranteed stability margin has been depleted. Thus, the voltage stability index $\Delta$ can be used to assess the stability margin.

At the steady state, the voltage stability index $\Delta$ is a load-weighted reactive power demand adopted to assess the power system's stability status and quantify the voltage stability margin. With the data provided by synchrophasor measurements, the steady-state voltage stability of each operation point is represented by $\Delta$.
However, in the broad field of STVS, the voltage stability index $\Delta$ is not applicable anymore to voltage dynamics. The STVS of the power system cannot be assessed simply by comparing it with the constant threshold as in the steady state. Nevertheless, the stress on each load changes continuously when the power system suffers from a large disturbance. Thus, there should be a time-dependent load-weighted reactive power demand for STVS assessment. 

Therefore, the voltage dynamic features $\mathscr{F}_{T}$ from $0$ to $T$ can be defined as,
\begin{equation}
\mathscr{F}_{T} \triangleq  (\Delta_{0}, \Delta_{1}, \cdots, \Delta_{t}, \cdots, \Delta_{T}),\quad  t = 0, 1, \cdots, T
\end{equation}
where $\Delta_{t}\triangleq \mathbf{L_{s}}^{-1}(\mathbf{q_{L}})_{t}$ is the snapshot of the voltage dynamic feature of STVS at $t$-time. $(\mathbf{q_{L}})_{t}$ is the real-time reactive power demands of loads. The voltage dynamic features $\mathscr{F}_{T}$ are constructed by stacking the snapshots $\Delta_{t}$ from $0$ to $T$, respectively, the beginning and end of the voltage trajectory $V_{t}$. The voltage dynamic features are the low-dimensional topology-aware feature for the DL-based STVS assessment. 

The topological information of power grid has a significant impact on the accuracy and efficiency of stability assessment. In the practical power system, the STVS problem easily spreads from vulnerable load buses. The feature extraction of the grid topology helps to assess the extent of disturbances propagation and capture the dynamic interactions between different nodes. The proposed feature construction method generates topology-aware voltage dynamic features with the load matrix $\mathbf{L_{s}}$. The susceptance matrix in the load matrix contains the topological information and power system parameters. The elements in the load matrix reflect the mutual connection relationship between the corresponding nodes. Constructed with the raw PMU data with the load matrix and real-time power demands of loads, the voltage dynamic features preserve the topological information of the power grid as well as the temporal voltage dynamics in the STVS assessment.

\section{DL-based STVS Assessment and Transfer Learning}
\subsection{Framework of DL-based STVS Assessment }
\begin{figure}[!t]
\centering
\includegraphics[width=3.5in]{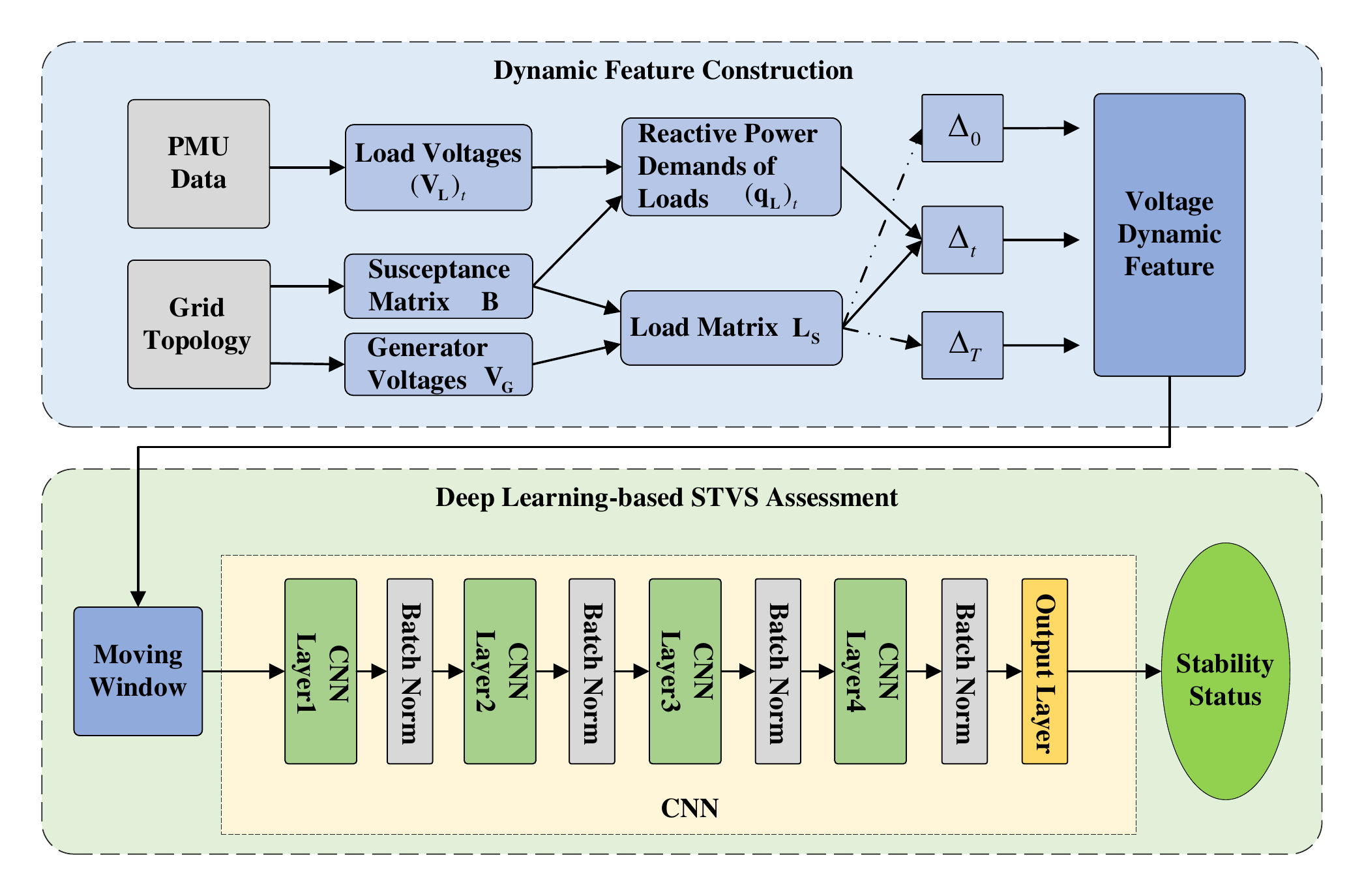}
\caption{Framework of the proposed feature construction methodology and DL-based STVS assessment model}
\label{struct}
\end{figure}

Based on the voltage dynamic features, the DL-based STVS assessment consists of two parts, the dynamic feature construction and the DL-based STVS assessment, as shown in Fig.~\ref{struct}. The voltage dynamic features, constructed with the grid topology and voltage dynamics, can enhance the knowledge transfer ability to the new grid topology. 
The STVS assessment is a binary classification problem. Therefore, DL algorithms are widely used for classification, particularly a CNN that is used for visual recognition and image classification. As shown in Fig.~\ref{struct}, the PMU data are converted into two-dimensional voltage dynamic features. Then, the voltage dynamic features are segmented by the moving windows as the input for the CNN network.

The CNN network consists of four CNN layers, four batch normalization layers, and one output layer. The output layer includes a fully-connected layer and a SoftMax layer. The classifier's output is a binary vector representing the predicted stability status of the power system. 

In the CNN layer, units are organized in the feature map, and each unit is connected to the local patch of the feature map of the previous layer through a filter blank. Mathematically, this transformation operation is a process of discrete convolution. The calculation of a CNN layer $\bm{H}_{1}$ is defined as:
\begin{equation}
\bm{h}_{1,\alpha}=f(\bm{x}*\bm{W}^{1,\alpha}+\bm{b}^{1,\alpha})
\end{equation}
where, $\bm{h}_{1,\alpha}$ is the output matrix of the $\alpha^{th}$ convolution of $\bm{H}_{1}$. $\bm{W}^{1,\alpha}$ and $\bm{b}^{1,\alpha}$ are the weight matrix and bias matrix of the $\alpha^{th}$ convolution of $\bm{H}_{1}$. $f$ is the activation function.

\subsection{Transfer Learning with Fine-tuning}
\begin{figure}[!t]
\centering
\includegraphics[width=3.5in]{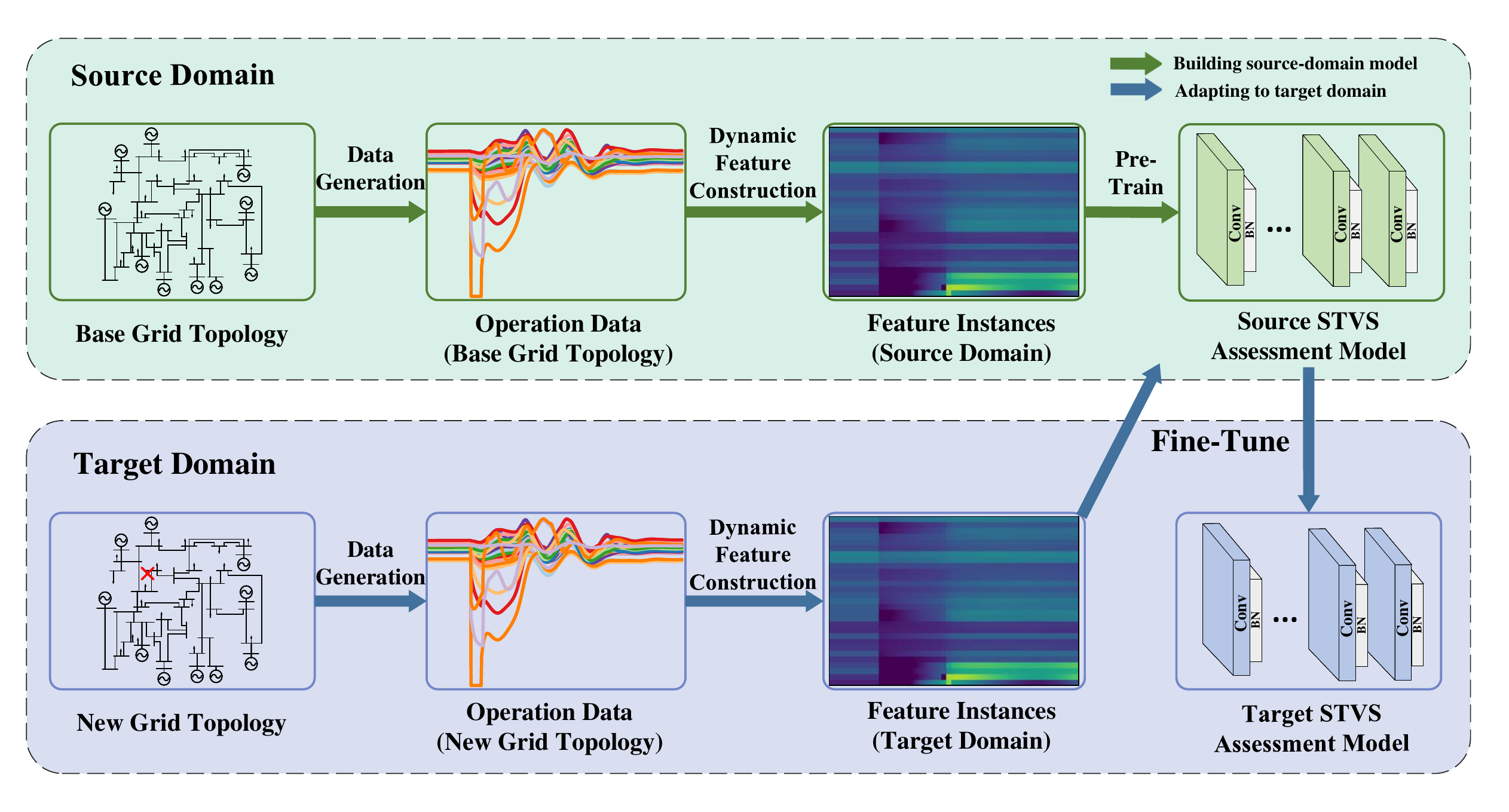}
\caption{Framework of fine-tuning of DL-based STVS assessment model}
\label{TL}
\end{figure}

The DL model, which is trained on one power system, could not be applied to other power systems with different grid topologies.  
The transfer learning is to realize the transfer of the knowledge of a trained model from a domain (the source domain) to another domain (the target domain). 
The power system with the particular grid topology is used as the source domain. The power systems with new grid topology are used as the target source.
The knowledge transfer ability can be strongly enhanced across different power grids with topology-aware dynamic features.
The framework of the knowledge transfer is shown in Fig.~\ref{TL}. 

Applying the pre-trained model to other power grids is equivalent to transferring the model trained in the source domain to the target domain. Moreover, by transfer learning, it is possible to apply the optimized STVS assessment model to power systems with different topologies without collecting a large amount of data for repetitive training. 
Fine-tuning is widely used for transfer learning\cite{TL2} to train further the pre-trained source domain model for the target domain. The STVS assessment model is pre-trained with dynamic features in the source domain. Then, after fine-tuning with the target domain's dynamic features, the pre-trained STVS assessment model from the source task is transferred to the target task. 

\section{Case Study}
\begin{figure}[!t]
\centering
\includegraphics[width=3.5in]{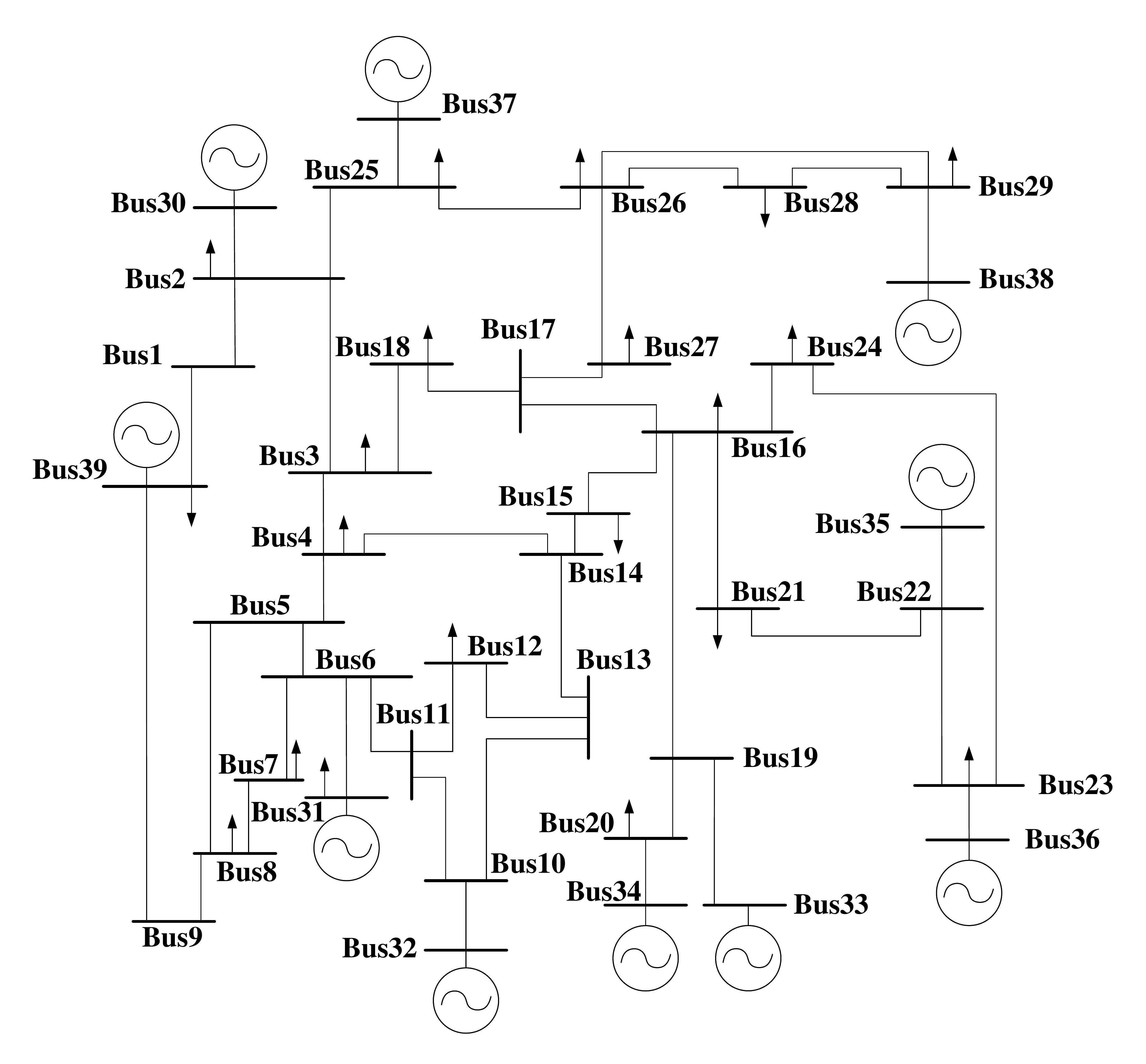}
\caption{New England 39-bus power system}
\label{IEEE39}
\end{figure}

The DL-based STVS assessment is tested in the New England 39-bus system shown in Fig.~\ref{IEEE39}.
The impact of load dynamics on the STVS performance of various power systems is examined following the short-circuit disturbances with different severity degrees. 

\begin{figure}[!t]
\centering
\includegraphics[width=3.5in]{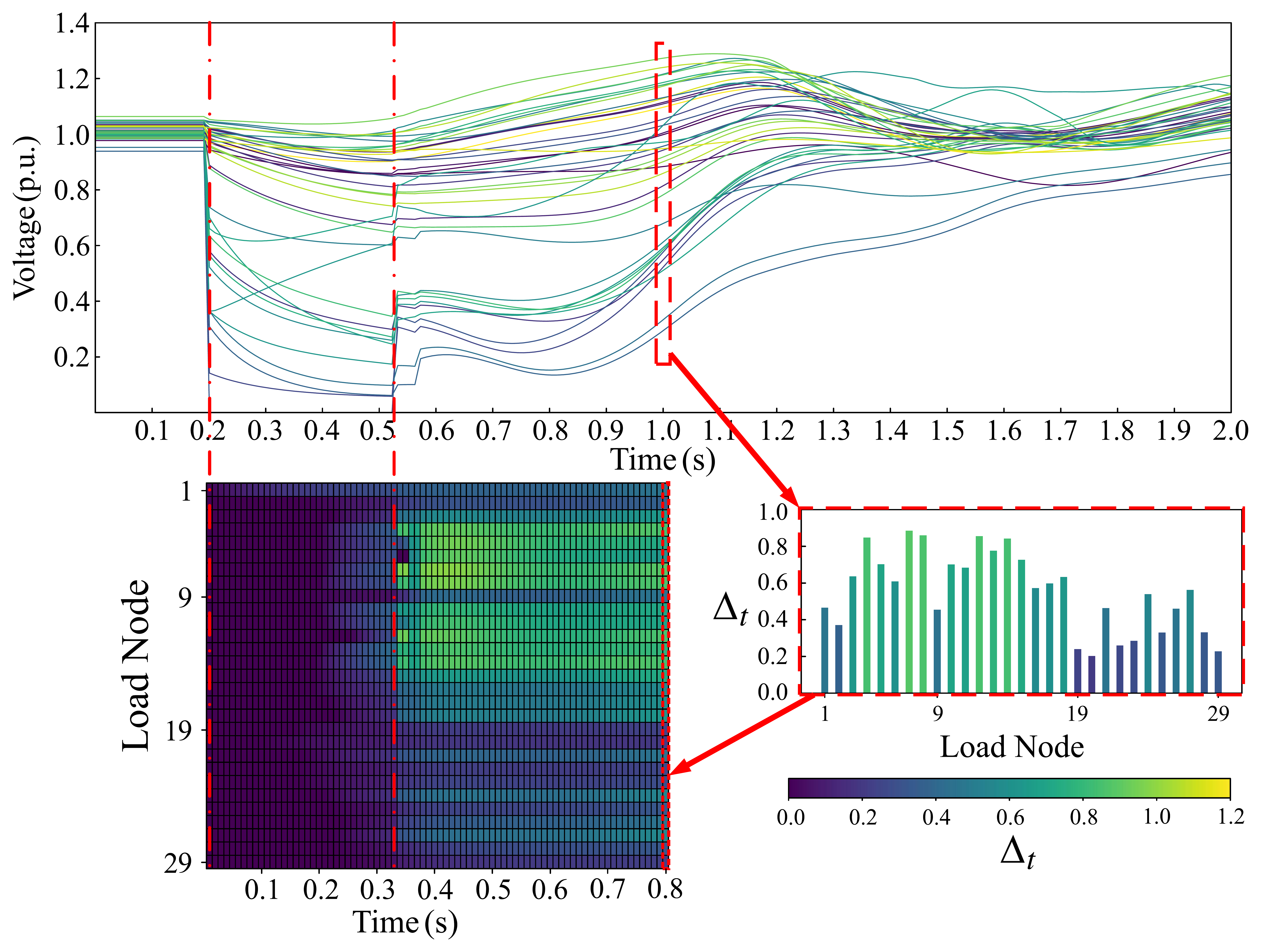}
\caption{The voltage trajectories and voltage dynamic features}
\label{method}
\end{figure}

The voltage trajectories are obtained by the TDS with Power System Toolbox\cite{PST}. The dynamic feature construction and STVS assessment are implemented in Python, and the framework of CNN is implemented in PyTorch. All the loads are modeled as composite loads with IMs and constant power loads to simulate the STVS dynamic. In a power system with $m$ loads, the voltage trajectories $V_{t}$ have the length of $T$. Given the time step of $dt$, the size of the voltage trajectory is $N=T/dt$. 
The voltage trajectory is separated into moving windows to construct the voltage dynamic features. The moving window has the $T_{w}$ length, and the window size is $n=T_{w}/dt$. 
The $\Delta_{t}$ is calculated for each load at each snapshot in the moving window. And the voltage dynamic features $\mathscr{F}_{T_{w}}$ are generated by stacking.
In Fig.~\ref{method}, the upper plot shows the voltage trajectory under the large disturbance, and the lower right plot shows the corresponding voltage dynamic feature of the voltage dynamics at $t$-time snapshot. The lower left plot shows the voltage dynamic features in a moving window.

\begin{figure}[!t]
\centering
\includegraphics[width=3.5in]{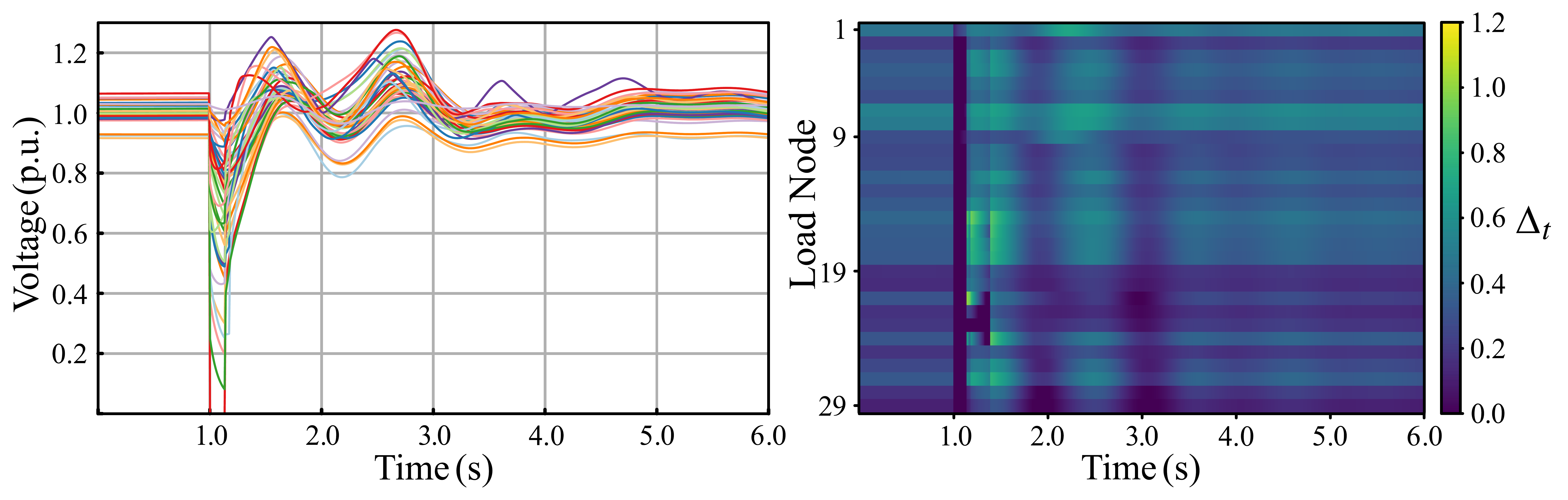}
\caption{Voltage dynamic features of stable status}
\label{sample1}
\end{figure}

\begin{figure}[!t]
\centering
\includegraphics[width=3.5in]{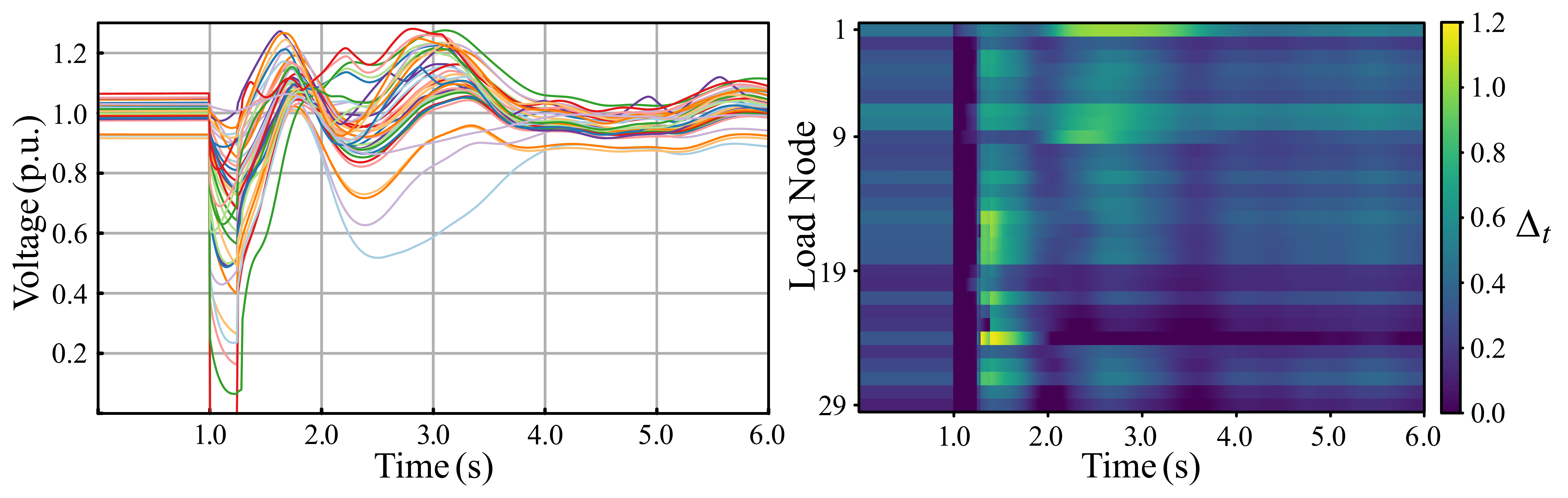}
\caption{Voltage dynamic features of unstable status}
\label{sample2}
\end{figure}

Figs.~\ref{sample1} and ~\ref{sample2} show the voltage trajectories and corresponding voltage dynamic features for the stable and unstable status. And the trajectories of voltage amplitude $V_{m}$ and voltage phase angle $\theta$ are obtained. The heatmap visualization of the voltage dynamic features clearly show that the stable and unstable status have different patterns. The pattern in the topology-aware voltage dynamic features enhances the ability of DL classification.

\subsection{Pre-trained DL Classifier}

The source-domain dataset is used to pre-train the DL classifier. In the source-domain dataset, varied operating conditions and contingencies, including the load level and fault clearing time, cover typical STVS scenarios. The source-domain dataset $G_{0}$ contain 100,000 samples.  The load level varies between 0.8 and 1.2 of its base values in data generation. Each sample has a three-phase fault with a random fault duration between $0.1$ and $0.4s$ and a randomly selected fault location. The label in the dataset is determined by whether the nodal voltage magnitude is lower than $0.8 p.u.$ for more than $1s$ in the simulation process~\cite{CS1}. To sufficiently describe the voltage dynamics by voltage dynamic features, the length of the moving window $T_{w}$ is set to $0.8s$. The decision of the beginning of the moving window depends on when the fault occurs. The source-domain dataset is split into the training and validation sets with 60000 and 20000 samples, respectively, via random sampling, and the testing dataset with 20000 samples. The proposed STVS assessment model is pre-trained by offline learning. 
The accuracy (ACC), Precision, Recall, and F1-score are used for the performance evaluation.

To assess the generalization ability of the proposed model, the cross-validation method is used. Cross-validation is one of the most popular data resampling methods to estimate the true prediction error of the models \cite{cross}. During the offline training of the STVS assessment model, the training, validation, and test dataset are generated by $k$-fold random subsampling, where $k=10$. Any set of the dataset is disjoint and in the same proportion. The performance is estimated as the average over all k-test datasets. This approach makes the evaluation results of the model more reliable.
The proposed STVS assessment model's accuracy reaches 99.99\%. Compared with the existing methods\cite{LR4,LR5,LR6}, the proposed model achieves the best overall STVS performance on the New England 39-bus system.

To examine the dependence of the pre-trained model on the dataset size, training datasets of different sizes, ranging from 1000 to 50000, are used for the training. When the training dataset has 10000 samples, the pre-trained model still reaches good accuracy at 99.63\%. However, when the size of the training dataset drops to 2000, the accuracy declines to 97.75\%. The sufficiently large training dataset guarantees the performance of the pre-trained model.

\begin{table}[htbp]
\centering
\caption{Performance on Different Size of Datasets}
\begin{tabular}{ccccccc}
	\toprule
	Size & ACC(\%) & Precision(\%) & Recall(\%) &  F1-score(\%) \\
	\midrule
	100000 & 99.99  & 99.99 & 99.99  & 99.99 \\
	50000 & 99.96  & 99.97  & 99.97  & 99.97 \\
	20000 & 99.84  & 99.87  & 99.92   & 99.89  \\
	10000 & 99.63  & 99.64  & 99.86 & 99.75  \\
	5000  & 98.85  & 99.18  & 99.30  & 99.24  \\
	2000  & 97.75  & 98.49  & 98.48  & 98.49  \\
	\bottomrule
\end{tabular}%
\label{size}%
\end{table}%

Samples in moving windows of different lengths from 0.1 to 0.8$s$ were used to compare the size effect on the performance of the pre-trained model. The models' performances trained with different lengths of moving windows are shown in Table~\ref{length}. When the length is not less than 0.4s, the accuracy of the models is not less than 99.61\%, which is still an outstanding result. Meanwhile, when the length is smaller, the models' performances decrease due to the insufficient temporal information of the sample. In this case, the length of the moving window $T_{w}$ is set to $0.8s$ in order to obtain better STVS assessment performance.

\begin{table}[htbp]
\centering
\caption{Size Effect of Moving Windows}
\begin{tabular}{ccccccc}
	\toprule
	Time Length & ACC(\%) & Precision(\%) & Recall(\%) &  F1-score(\%) \\
	\midrule
	0.8 & 99.99  & 99.99 & 99.99  & 99.99 \\
	0.7 & 99.95  & 99.97  & 99.97  & 99.97 \\
	0.6 & 99.85  & 99.85  & 99.95   & 99.90  \\
	0.5 & 99.79  & 99.76  & 99.97 & 99.86  \\
	0.4  & 99.61  & 99.71  & 99.79  & 99.75  \\
	0.3  & 97.57  & 98.41  & 98.44  & 98.42  \\
	0.2  & 93.64  & 94.77  & 97.13  & 95.93 \\
	0.1  & 91.07  & 92.79  & 95.89  & 94.31 \\
	\bottomrule
\end{tabular}%
\label{length}%
\end{table}%

Due to possible errors caused by phasor measurement, the performance of the proposed assessment model was further tested on voltage data with PMU error. Based on the previous experiment, measurement errors were added to the voltage trajectory dataset. In particular, errors were simulated by white noise following Gaussian distribution $N(0,\sigma)$ to voltage magnitudes $(\sigma=0.3p.u)$ and angles $(\sigma=1.5^{\circ})$ \cite{CS3}. Accordingly, the voltage dynamic features were calculated using voltage data with PMU error. 

Compared with the error-free performances, the proposed model's STVS assessment with PMU errors still has excellent performances, with accuracy at 99.52\%, precision at 99.70\%, recall at 99.69\%, and F1-score at 99.70\%. The results imply that the proposed assessment model remains robust to the PMU error. This effect benefits from the proposed feature construction method that extracts voltage dynamic voltage features from voltage trajectories and acts as the filter of noise. Even if noises contaminate the acquired voltage trajectories, the patterns in voltage dynamic features are still preserved behind them.

\subsection{Transfer Learning Classifier}

Unexpected overloading operation or maintenance scheduling processes of transmission lines will cause topology changes in the power system. In the experiment, the single-line diagram of the standard New England 39-bus system is adjusted for different scenarios of topology changes for data generation. The new power grids are selected to evaluate the transfer learning performance. The target-domain datasets of transfer learning are generated based on two power grid topology change scenarios. The new two scenarios are as follows:

Scenario A: One edge of the standard New England 39-bus system is disconnected, and the load level is unchanged. Six target domain datasets $G_{1}$ to $G_{6}$ are generated.

Scenario B: Two edges of the standard New England 39-bus system are disconnected, and the load level is unchanged. Six target domain datasets $G_{7}$ to $G_{12}$ are generated.

The detailed setup of target power grid topology changes is shown in Table~\ref{TL1}.

\begin{table}[htbp]
\centering
\caption{Setup of Topology Changes}
\begin{tabular}{cccc}
	\toprule
	\multicolumn{1}{c}{Scenario}  & \multicolumn{1}{c}{Dataset} & \multicolumn{1}{c}{Line(s) Disconnected} & \multicolumn{1}{c}{Dataset size} \\
	\midrule
	\multicolumn{1}{c}{\multirow{6}[1]{*}{\makecell[c]{A}}} 
	& $G_{1}$ & Line 2-3 &  5000 \\
	& $G_{2}$ & Line 5-8 &  5000 \\
	& $G_{3}$ & Line 14-15 &  5000 \\
	& $G_{4}$ & Line 4-14 &  5000 \\
	& $G_{5}$ & Line 16-24 &  5000 \\
	& $G_{6}$ & Line 17-18 &  5000 \\
	\midrule
	\multicolumn{1}{c}{\multirow{6}[1]{*}{\makecell[c]{B}}} 
	& $G_{7}$ & Lines 2-3 \& 5-8 &  5000 \\
	& $G_{8}$ & Lines 4-14 \& 14-15 &  5000 \\
	& $G_{9}$ & Lines 16-24 \& 17-18 &  5000 \\
	& $G_{10}$ & Lines 14-15 \& 16-24 &  5000 \\
	& $G_{11}$ & Lines 4-14 \& 17-18 &  5000 \\
	& $G_{12}$ & Lines 14-15 \& 17-18 &  5000 \\
	\bottomrule
\end{tabular}%
\label{TL1}%
\end{table}%

Firstly, without further processing, the pre-trained classification model is directly applied to the target-domain datasets $G_{1}$ to $G_{12}$. The direct transfer learning performance is shown in Table~\ref{FT}. Direct transfer learning has relatively good performance on the target datasets. Dataset $G_{1}$ to $G_{6}$ reach the accuracy from 94.60\% to 98.64\% whereas $G_{7}$ to $G_{12}$ from 88.76\% to 94.18\%. 

However, fine-tuning the pre-trained model with samples in the target-domain dataset can enhance knowledge transfer performance. In the 12 target-domain datasets, $G_{1}$ to $G_{12}$, 1000 samples are used for the fine-tuning. The test results on the 5000 testing dataset are shown in Table~\ref{TL1}.

In Table~\ref{FT}, it can be seen that the transfer learning STVS assessment model achieves better performance on the 12 target-domain datasets after fine-tuning. The best accuracy can reach 99.68\%. With more edges disconnected in the new grid topology, the transfer performance declines due to the more significant power grid topology change. However, the lowest accuracy still reaches 97.1\%. It shows that the pre-trained model has a superior transfer ability to power systems with new grid topology.

\begin{table}[htbp]
\centering
\caption{Performance of Transfer Learning}
\begin{tabular}{cccccc}
	\toprule
	\multicolumn{1}{c}{ }  & \multicolumn{1}{c}{Dataset} & \multicolumn{1}{c}{ACC(\%)} & \multicolumn{1}{c}{Precision(\%)} & \multicolumn{1}{c}{Recall(\%)}  & \multicolumn{1}{c}{F1-score(\%)} \\
	\midrule
	\multicolumn{1}{c}{\multirow{12}[1]{*}{\makecell[c]{Direct\\transfer\\learning}}} 
	& $G_{1}$ & 94.72  & 94.22  & 99.33  & 96.71  \\
	& $G_{2}$ & 94.60  & 93.69  & 99.58  & 96.54 \\
	& $G_{3}$ & 97.20  & 98.05  & 98.35  & 98.20  \\
	& $G_{4}$ & 98.64  & 98.92  & 99.41  & 99.17  \\
	& $G_{5}$ & 97.16  & 97.59  & 99.71  & 98.86  \\
	& $G_{6}$ & 98.16  & 97.89  & 99.85  & 98.86  \\
	\cmidrule(lr){2-6}
	& $G_{7}$ & 92.76  & 91.77  & 99.31  & 95.39  \\
	& $G_{8}$ & 93.49  & 93.49  & 98.03  & 95.71  \\
	& $G_{9}$ & 88.76  & 89.22  & 97.51  & 93.18  \\
	& $G_{10}$ & 94.18  & 94.49  & 98.04  & 96.23  \\
	& $G_{11}$ & 94.07  & 92.93  & 99.68  & 96.19  \\
	& $G_{12}$ & 93.36  & 92.95  & 97.79  & 95.31  \\
	\midrule
	\multicolumn{1}{c}{\multirow{12}[1]{*}{\makecell[c]{Fine-\\tuning}}} 
	& $G_{1}$ & 99.56  & 99.74  & 99.69  & 99.72  \\
	& $G_{2}$ & 99.48  & 99.37  & 99.95  & 99.66  \\
	& $G_{3}$ & 99.16  & 99.28  & 99.64  & 99.46  \\
	& $G_{4}$ & 99.68  & 99.61  & 100.00  & 99.80  \\
	& $G_{5}$ & 99.56  & 99.76  & 99.71  & 99.74  \\
	& $G_{6}$ & 99.28  & 99.21  & 99.90  & 99.55  \\
	\cmidrule(lr){2-6}
	& $G_{7}$ & 99.36  & 99.79  & 99.36  & 99.58  \\
	& $G_{8}$ & 99.62  & 100.00  & 99.49  & 99.75 \\
	& $G_{9}$ & 97.84  & 98.63  & 98.63  & 98.63  \\
	& $G_{10}$ & 98.45  & 98.89  & 99.06  & 98.98  \\
	& $G_{11}$ & 98.68  & 99.36  & 98.88  & 99.12  \\
	& $G_{12}$ & 98.11  & 97.99  & 99.29  & 98.64  \\
	\bottomrule
\end{tabular}%
\label{FT}%
\end{table}%
\section{Conclusion}
The DL-based power system STVS assessment model is proposed with topology-aware and physics-informed feature engineering and knowledge transfer for power grid topology change. After research, it is found that more than the existing methods are needed to improve the model to adapt to the new power grid topology. With transfer learning as a promising method for this problem, this paper proposed a novel dynamic feature construction method. It used fine-tuning to adapt the pre-trained model to a new grid topology. The feature construction method generates topology-aware voltage dynamic features from PMU data, with CNN as a classifier to fit the relationship between the voltage dynamic features and STVS status. Case studies were carried out on the New England 39-bus system, and the results show that the proposed assessment achieves good performances and adaptability on the robustness to PMU error. Moreover, the pre-trained model can significantly ensure knowledge transfer ability to new grid topology.
It is very promising to further explore the applications of the STVS assessment with voltage dynamic features on voltage stability emergency control.

	\bibliographystyle{IEEEtran}
	\bibliography{References}
	
	
\end{document}